\documentclass[a4paper]{PoS}
\usepackage{amsmath}
\usepackage{amssymb}
\usepackage{subfigure}

\allowdisplaybreaks[1]

\title{Probing gaugino coannihilation with displaced vertex searches}

\ShortTitle{Probing gaugino coannihilation with displaced vertex
searches}
\author{\speaker{Natsumi Nagata}%
        \thanks{Based on work in collaboration with Hidetoshi Otono and
        Satoshi Shirai \cite{Nagata:2015hha, Nagata:2015pra}.}\\
       William I. Fine Theoretical Physics Institute, School of Physics
        and Astronomy, \\University of Minnesota, Minneapolis, Minnesota
        55455, USA\\
       E-mail: \email{nagat006@umn.edu}}

\abstract{Bino-like dark matter in supersymmetric theories tends to
suffer from over-production. This problem can be evaded if there is
another gaugino which has a mass close to the bino mass so that it
enhances the annihilation rate of bino dark matter through
coannihilation. We study such possibilities in high-scale supersymmetric
models, and show that searches for long-lived particles with displaced
vertices can test the bino-gaugino coannihilation scenario in the
forthcoming LHC experiments. }

\FullConference{The 11th International Workshop Dark Side of the Universe 2015\\
		14--18 December 2015\\
		Kyoto, Japan}

\begin{document}

\section{Introduction}

Despite plenty of evidence for the presence of dark matter (DM), we
have almost no idea what it is. Weakly-interacting massive particles
(WIMPs) are the most promising class of candidates for DM, since the
observed DM density can naturally be explained by their thermal relic
abundance. Many beyond-the-Standard-Model theories predict WIMPs as
their inevitable ingredients. Among them, the minimal supersymmetric
Standard Model (MSSM) offers a very attractive candidate for WIMP DM:
the lightest neutralino, which is an electrically neutral and colorless
fermion composed of a linear combination of bino
$\widetilde{B}$ (the superpartner of the U(1)$_Y$ gauge field), the
neutral wino $\widetilde{W}_3$ (winos $\widetilde{W}_a$, $a=1,2,3$, are the
superpartners of the SU(2)$_L$ gauge bosons), and higgsinos $H_u$ and
$H_d$ (the superpartners of the Higgs boson). The stability of the
neutralino DM is assured by the $R$-parity if it is the lightest
supersymmetric (SUSY) particle (LSP) \cite{Goldberg:1983nd}.

On the other hand, the recent LHC results have imposed stringent
constraints on TeV-scale physics. Regarding SUSY models,
multi-jets plus missing energy searches restrict masses of SUSY
particles, especially those of squarks and gluino \cite{Aad:2015iea}. In
addition, the mass of the Standard Model (SM) Higgs boson, $m_h \simeq 125$~GeV
\cite{Aad:2015zhl}, implies that SUSY partners are heavy, since large
quantum corrections to the Higgs mass by heavy stops are required in
order to achieve the observed Higgs mass in the MSSM. Consequently, the
early LHC results might indicate that the SUSY scale is much higher than
the electroweak scale, contrary to the expectation before the LHC run.

Considering this situation, the questions we would like to address here are
the following. Is the neutralino DM still promising? Can we probe the
neutralino DM in future experiments? To answer these questions, it is
important to list up the present allowed parameter regions for the
neutralino DM, and study their phenomenological consequences.

The constrained MSSM (CMSSM) is a widely studied benchmark scenario for
SUSY models. Recently, a parameter scan in the CMSSM
was performed in Ref.~\cite{Bagnaschi:2015eha}. This result shows that
the correct DM relic abundance can be obtained only in specific
parameter regions (see Ref.~\cite{Olive:2016efh} for more detailed
discussions). Most of the parameter points will be probed in future LHC
experiments, DM searches, or proton decay experiments
\cite{Bagnaschi:2015eha, Olive:2016efh, Ellis:2015rya}.  

On the other hand, SUSY models with a high SUSY-breaking scale,
as high as $100$--1000~TeV \cite{Wells:2003tf}, have attracted wide
attention recently, especially after the
Higgs discovery \cite{Hall:2011jd}. These models are
based on the assumption that there is no singlet SUSY-breaking field in
the SUSY breaking sector, which couples to the
MSSM sector via a generic form of the K\"ahler potential. In this case,
all of the scalar particles except the SM Higgs boson have masses
similar to the gravitino mass, $m_{3/2}$. The supersymmetric higgsino
mass $\mu$ is also expected to be around $m_{3/2}$, though it depends on
models. In what follows, we assume that a typical scale of soft masses,
$\widetilde{m}$ and the higgsino mass $\mu$ is ${\cal
O}(m_{3/2})$.\footnote{In models where $\mu \ll m_{3/2}$, which can be
achieved by means of a certain symmetry such as the Peccei--Quinn symmetry
\cite{Peccei:1977hh}, the neutral higgsino can be a viable dark matter
candidate. See Refs.~\cite{Bae:2014yta, Evans:2014pxa} for concrete
examples for the realization of the higgsino LSP. Regarding the
phenomenology of the higgsino DM scenario, see
Ref.~\cite{Nagata:2014wma} and references therein. } 
The gaugino masses are, on the other hand, induced only by quantum
corrections, and thus suppressed by a loop factor compared with
$m_{3/2}$. Such quantum effects include the anomaly mediation
\cite{Randall:1998uk, Giudice:1998xp} and threshold corrections by
heavy Higgs bosons \cite{Giudice:1998xp, Pierce:1996zz}, extra
vector-like multiplets whose masses are ${\cal O}(m_{3/2})$
\cite{Pomarol:1999ie}, or particles appearing around the grand
unification scale \cite{Hisano:1993zu}. We therefore expect that
the LSP is one of the gauginos, though their
mass spectrum depends on sizes of these quantum corrections. 

In this case, there are two DM candidates; one is the neutral wino
and the other is bino. The wino DM is known to be a good DM candidate. Its
thermal relic abundance agrees to the observed DM density if the wino
mass is $\simeq 3$~TeV \cite{Hisano:2006nn}. This DM can be probed in
both direct \cite{Hisano:2015rsa} and indirect
\cite{Bhattacherjee:2014dya} DM searches. 
On the other hand, the bino DM often suffers from over production due to
its small annihilation cross section. Thus, some additional mechanisms
are required to reduce its thermal relic. In the present scenario, we
can use coannihilation \cite{Griest:1990kh} with either gluinos
\cite{Profumo:2004wk,
Harigaya:2014dwa, Ellis:2015vaa, Nagata:2015hha} or winos
\cite{Baer:2005jq,  Harigaya:2014dwa,
Nagata:2015pra}. We discuss these two scenarios in this article.

Both of these scenarios require the lightest and the next-to-lightest
particles to be highly degenerate in mass, which makes it
difficult to probe these scenarios with traditional LHC search methods,
since decay products of the next-to-lightest SUSY particle (NLSP) tend
to be too soft. Furthermore, it is also hard to detect the bino DM in DM
searches since its interactions are extremely weak. It turns out,
however, that in both of these scenarios the NLSP becomes long-lived,
which results in specific signatures at collider experiments. Thanks to
this feature, we can probe these coannihilation scenarios using
displaced vertex (DV) searches \cite{Nagata:2015hha, Nagata:2015pra,
Otono:2015jta}, as we will see below.

\section{Bino-gluino coannihilation}

\begin{figure}[t]
\centering
\includegraphics[clip, width = 0.6 \textwidth]{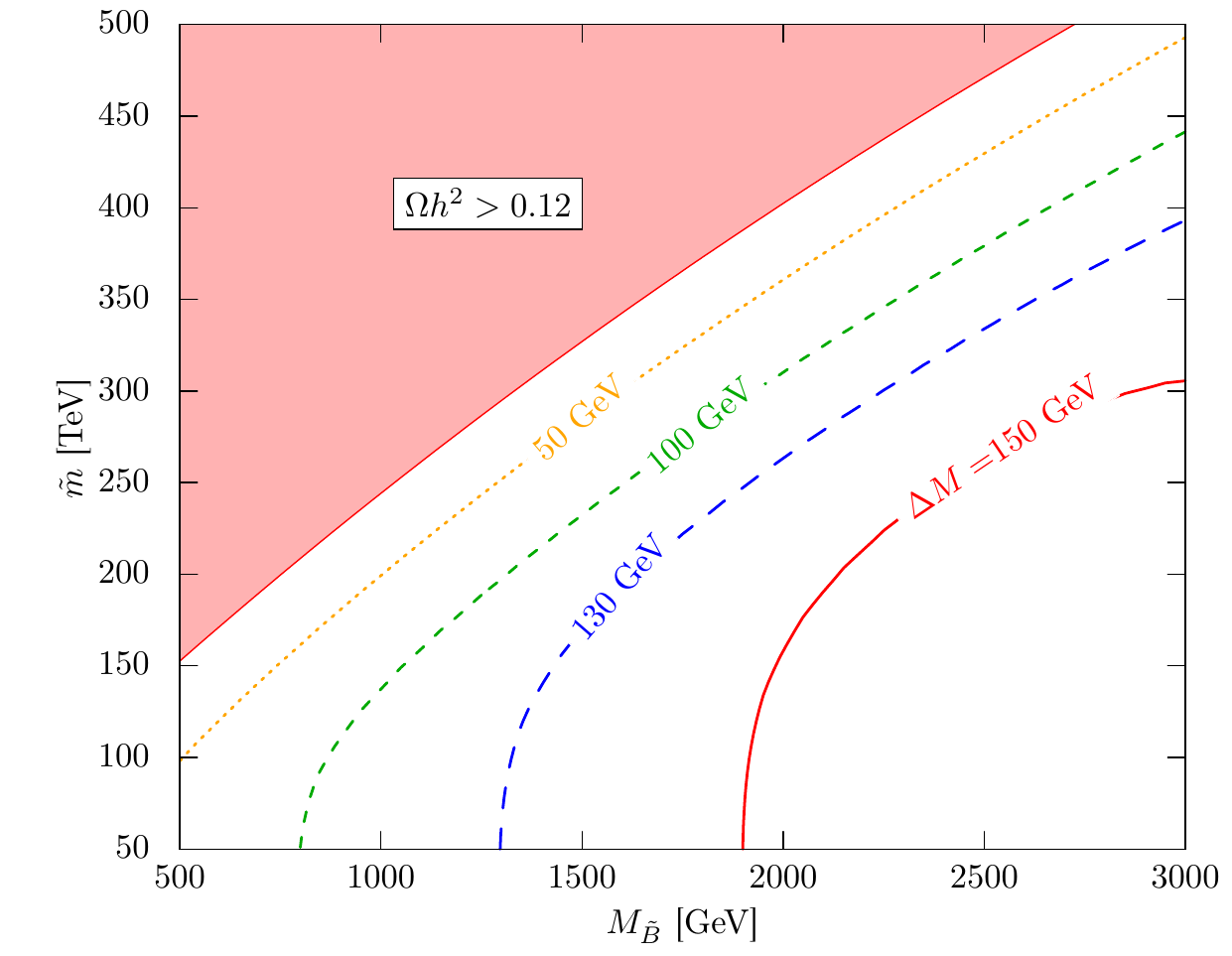}
\caption{Contours for the mass difference $\Delta M$ which achieves
 $\Omega_{\rm DM}h^2 = 0.12$.}
\label{fig:sq}
\end{figure}

First we
consider the case in which gluino is degenerate with bino in mass so
that the relic abundance of the bino DM can be sufficiently reduced
because of the coannihilation mechanism. In order for the coannihilation
to work effectively, however, chemical equilibrium between bino and
gluino should be maintained until the time when the bino DM is decoupled
from thermal bath \cite{Ellis:2015vaa, Nagata:2015hha}. This requires
the transition rate of bino into gluino, $\Gamma (\widetilde{B}q \to
\widetilde{g} q)$, be much larger than the Hubble expansion rate,
$H$. This transition is induced by the squark exchange diagrams and its
amplitude is suppressed by $\widetilde{m}^{-2}$. Therefore, 
this requirement 
gives an upper bound on $\widetilde{m}$. In
Fig.~\ref{fig:sq}, we show contour lines for the required mass
difference between bino and gluino, $\Delta M \equiv M_{\widetilde{g}} -
M_{\widetilde{B}}$ ($M_{\widetilde{g}}$ and $M_{\widetilde{B}}$ are the
masses of gluino and bino, respectively), to explain
the observed DM density $\Omega_{\rm DM}h^2 = 0.12$ with the thermal
relic of the bino LSP. In the pink shaded area, bino is over produced
because the transition rate $\Gamma (\widetilde{B}q \to \widetilde{g}q)$
is too small. From this figure we find that the bino-gluino
coannihilation scenario requires $\Delta M \simeq 100$~GeV. 

With such a small $\Delta M$, gluinos tend to be long-lived. Gluinos
decay into SM quarks and a bino again via the squark exchange. The
gluino decay length is roughly given by 
\begin{equation}
 c\tau_{\widetilde{g}} \simeq
\left(\frac{\Delta M}{100~{\rm GeV}}\right)^{-5}
\left(\frac{\widetilde{m}}{100~{\rm TeV}}\right)^4 ~{\rm cm} ~,
\end{equation}
and thus in the bino-gluino coannihilation scenario we expect ${\cal
O}(1)$~cm decay length, which can be a good target of the DV searches as
we will discuss below.

\begin{figure}[t]
\centering
\includegraphics[clip, width = 0.6 \textwidth]{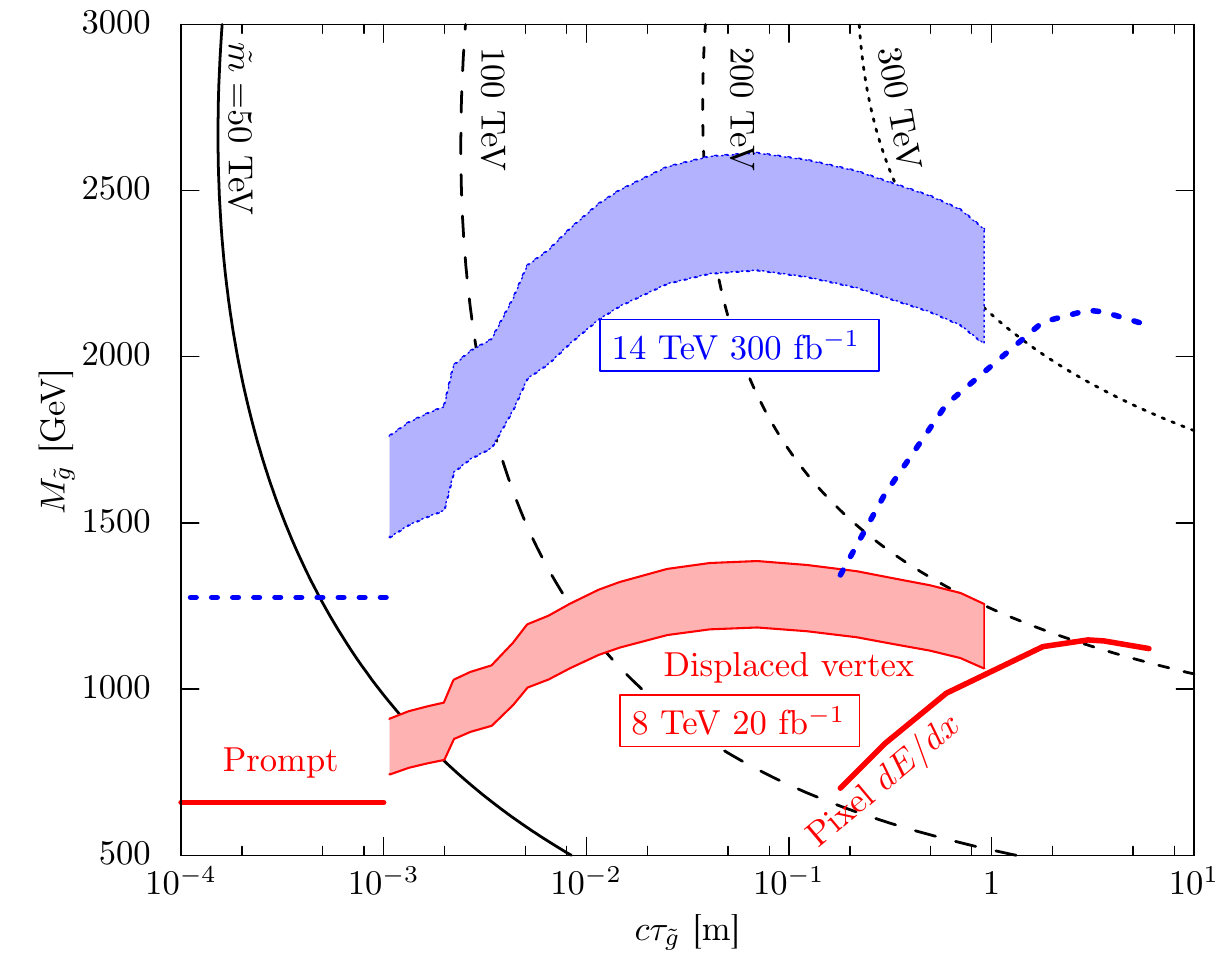}
\caption{Current limits (red solid) and future prospects
 (blue dashed) of long-lived gluino searches \cite{Nagata:2015hha}. }
\label{fig:prospect}
\end{figure}

The ATLAS collaboration has searched for DVs in the inner detector
\cite{Aad:2015rba}, and thus this
search is sensitive to decay lengths of $1{\rm mm} \lesssim c\tau
\lesssim 1$~m. Among the several search categories in
Ref.~\cite{Aad:2015rba}, the DV plus missing transverse energy ($E_{\rm
T}^{\rm miss}$) search is most sensitive to the present setup. Using the
data collected at the LHC 8-TeV run with an integrated luminosity of
$20.3$~fb$^{-1}$, the ATLAS collaboration has searched for such events,
and has observed no signal. This result then gives a lower limit on
$M_{\widetilde{g}}$ \cite{Aad:2015rba}; for instance, if the neutralino
LSP mass is 100~GeV, a lower limit on $M_{\widetilde{g}}$ is given as
$M_{\widetilde{g}} \gtrsim 1.5$~TeV for $c\tau_{\widetilde{g}} \simeq
10$~cm. We however note that we cannot directly apply this result to the
bino-gluino coannihilation case, since in this case we need to take into
account the fact that $\Delta M $ is as small as $100$~GeV. To that end, we
simulate the reduction of trigger efficiency due to the small $\Delta
M$ by using {\tt HERWIG6} \cite{Corcella:2000bw} and {\tt AcerDET}
\cite{RichterWas:2002ch}, and reinterpret the ATLAS result. Here, we
require $E_{\rm T}^{\rm miss} > 100$~GeV for the 8-TeV run. For the
estimation of the future prospects, we assume the 14~TeV run with an
integrated luminosity of 300~fb$^{-1}$, and require $E_{\rm T}^{\rm
miss} > 200$~GeV. 
In Fig.~\ref{fig:prospect} we show the current constraints and future
prospects for the long-lived gluino searches in red solid and blue
dashed lines, respectively. For $c\tau_{\widetilde{g}}< 1$~mm, we use
the ordinary limit from the gluino search
based on jets plus $E_{\rm T}^{\rm miss}$ only
\cite{Chatrchyan:2014lfa}. For $10^{-3}~\text{m} < c\tau_{\widetilde{g}}
< 1$~m, we show our estimate based on the
ATLAS DV search \cite{Aad:2015rba}, while for
$c\tau_{\widetilde{g}} \gtrsim 10^{-1}$~m, we plot the limit from the
$R$-hadron search \cite{Aad:2015qfa}. To estimate the uncertainty of our
evaluation, we vary reconstruction efficiency for DVs between 20--100\%
of that in Ref.~\cite{Aad:2015rba}, which is shown in the bands in the
figure. We also show the
black contour lines along which the observed DM density is obtained with
several values of $\widetilde{m}$. This figure shows
that a wide region of parameter space favored by the bino-gluino
coannihilation scenario can be probed with the DV search. In particular,
its sensitivity is found to be better than those of the ordinary
gluino searches based on multi-jets plus $E_{\rm T}^{\rm miss}$.

\section{Bino-wino coannihilation}

\begin{figure}[t]
\centering
\includegraphics[clip, width = 0.6 \textwidth]{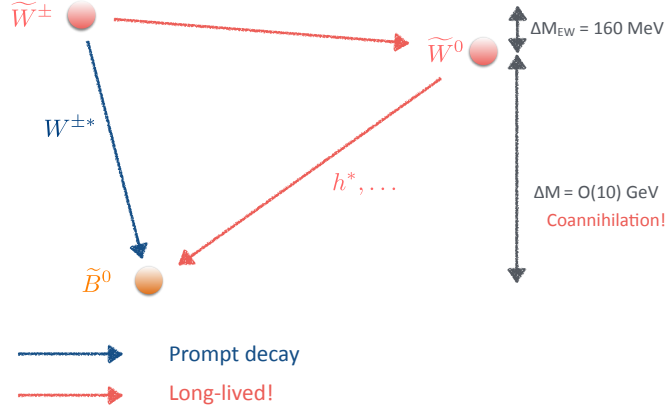}
\caption{Mass spectrum and decay chains of the bino-wino system.}
\label{fig:bwfig}
\end{figure}

\begin{figure}[t]
\begin{center}
\subfigure[Decay length]
{\includegraphics[clip, width = 0.49 \textwidth]{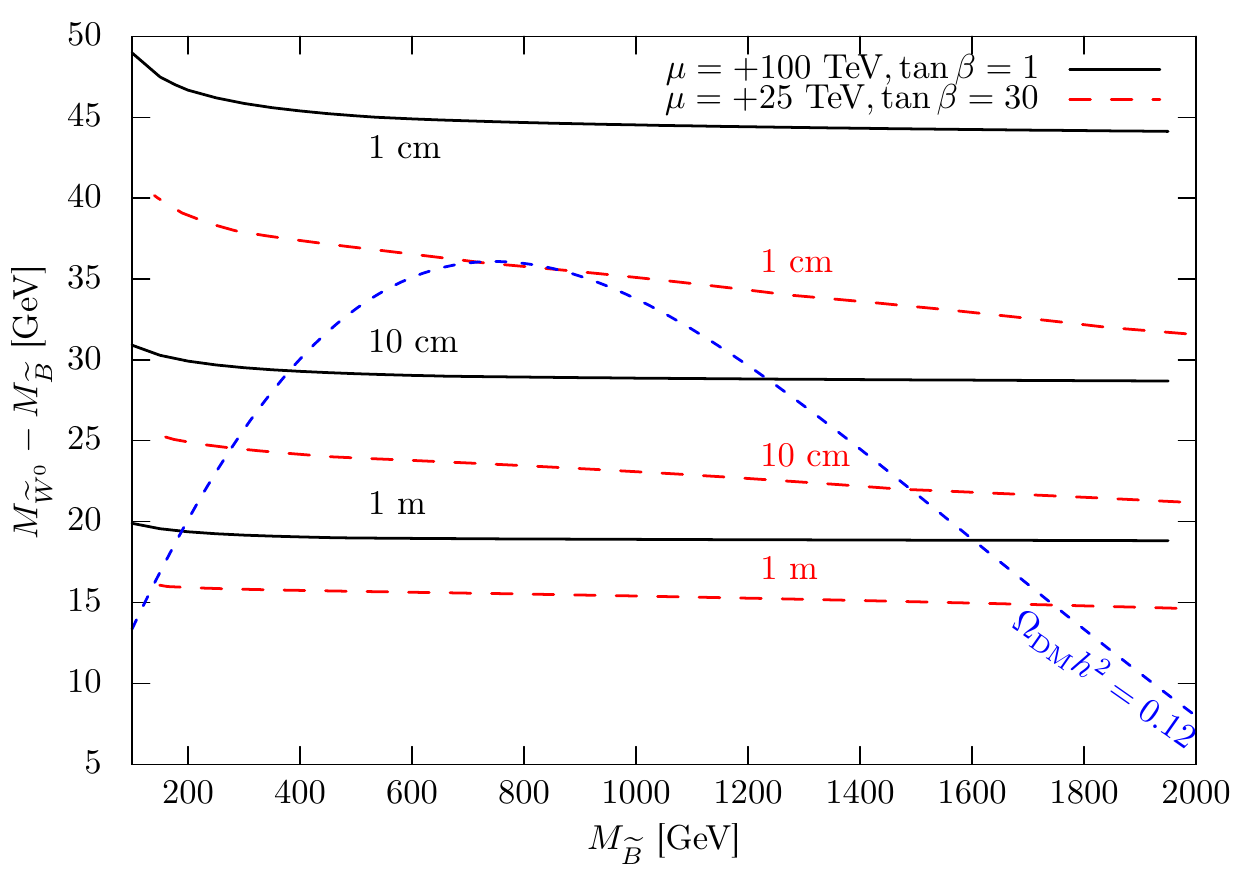}
\label{fig:decaylength}}
\subfigure[Constraints and prospects]
{\includegraphics[clip, width = 0.49 \textwidth]{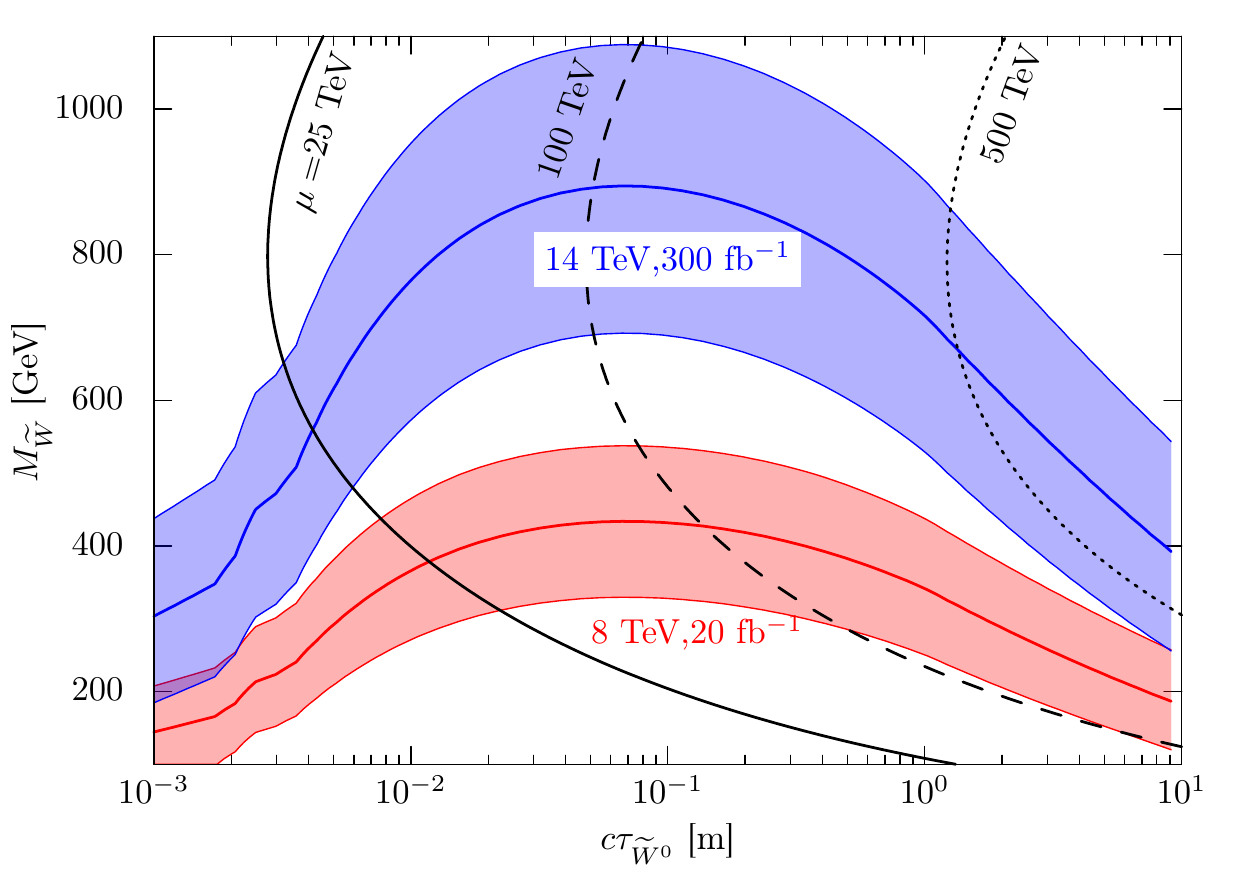}
\label{fig:cons}}
\caption{(a) The decay length of the neutral wino. (b) The present limit
 (red) and future prospects (blue) of the long-lived
 neutral wino search \cite{Nagata:2015pra}. 
 }
\label{fig:neutwino}
\end{center}
\end{figure}

Next, we discuss the case of the bino-wino coannihilation scenario
\cite{Nagata:2015pra}. It turns out that the NLSP again tends to be
long-lived over the parameter region favored by the bino-wino
coannihilation scenario. The relevant mass spectrum is
shown in Fig.~\ref{fig:bwfig}. As we will see below, the bino-wino
coannihilation scenario requires the bino-wino mass difference, $\Delta
M$, to be ${\cal O}(10)$~GeV. A neutral wino can decay into the bino LSP
via mixing with higgsinos. Therefore, the decay rate is suppressed if
$\mu$ is very large. It is found that in
this case its dominant decay mode is induced by an off-shell Higgs boson
emission. Thus, the amplitude is also suppressed by small Yukawa
couplings. Notice that the $Z$
boson exchange process is sub-dominant in this setup. The decoupling
features of these decay amplitudes with respect to large $\mu$ is studied in
Ref.~\cite{Nagata:2015pra} using the effective field theoretical
approach. Moreover, there is an additional suppression factor due to the
small mass difference $\Delta M$. For these reasons, the decay length of
the neutral wino NLSP becomes about $1$~cm--1~m, which can again be
probed with DV searches \cite{Nagata:2015pra, Rolbiecki:2015gsa}. In
Fig.~\ref{fig:decaylength}, we show the 
decay length of the neutral wino in the black solid (red dashed)
contour lines in the $M_{\widetilde{B}}$--$\Delta M$ plane, where we set
$\mu = 100$~TeV and $\tan \beta = 1$ ($\mu = 25$~TeV and $\tan \beta =
30$). The blue dashed line corresponds to the parameter points where the
observed DM density is explained by the thermal relic of the bino
LSP. As can be seen, the neutral wino has a decay length of $\gtrsim
1$~cm over the parameter region motivated by the bino-wino coannihilation
scenario. On the other hand, a charged wino, which has a slightly larger
mass than a neutral wino because of the radiative corrections, decays
into bino promptly through an off-shell emission of the $W$ boson as
long as $|\mu |<{\cal O}(10^4)$~TeV.

Similarly to the previous case, we reinterpret the ATLAS DV search
result \cite{Aad:2015rba} to study the current constraints and future
prospects of the bino-wino coannihilation scenario. 
For this purpose, we use the program packages {\tt
Madgraph5} \cite{Alwall:2011uj}, {\tt Pythia6} \cite{Sjostrand:2006za},
and {\tt Delphes3} \cite{deFavereau:2013fsa} for the estimation of the
acceptance rate, while we use {\tt Prospino2} \cite{Beenakker:1996ed}
for the computation of the cross sections. In this analysis, we have
dropped the $P_{\rm T}$ condition for DVs adopted in
Ref.~\cite{Aad:2015rba}, to optimize this search for the bino-wino
coannihilation scenario. In this case, a large $E^{\rm miss}_{\rm T}$
results from the back reaction of initial state radiations. We then
obtain the current constraints and future prospects of the long-lived
neutral wino search, which is shown in Fig.~\ref{fig:cons} in red
and blue lines, respectively. Here, we set $\Delta M =30$~GeV
and $\tan\beta =2$. The bands represent the uncertainties of our
estimation, where we vary the acceptance rate by a factor of three. 
The black contour lines correspond to the parameter
points where the correct DM density is obtained for $\mu = 25$, 100, and
500 TeV, respectively (from left to right). 
This figure shows that a wino with a mass of 400 (800) GeV can be probed
at the 8 (14) TeV LHC in the case of the bino-wino coannihilation
scenario.

\section{Summary}

In this article, we discuss possibilities of bino DM in the high-scale
SUSY scenario. In this case, the observed DM density can be explained if
gluino or wino is degenerate with the bino LSP in mass. Because of the
mass degeneracy as well as large masses of the exchanged particles in
the decay processes, the NLSP tends to have long lifetime, which can be
probed with DV searches. For the bino-gluino coannihilation scenario,
the 14~TeV LHC run can reach a gluino mass of $\sim 2$~TeV, while for
the bino-wino coannihilation scenario, the expected reach for the wino
mass is $\sim 800$~GeV.

\acknowledgments
The author thanks H. Otono and S. Shirai for collaboration. 
This work was supported by the DOE grant DE-SC0011842 at the University
of Minnesota. 



\end{document}